\documentstyle[prl,aps,multicol]{revtex}

\draft

\author{M. A. Nielsen\thanks{mnielsen@theory.caltech.edu}}
\title{Continuity bounds for entanglement}
\address{Physics MC 12-33, California Institute of Technology,
Pasadena CA 91125}
\date{\today}

\newcommand{\la}{\langle}
\newcommand{\ra}{\rangle}
\newcommand{\bea}{\begin{eqnarray}}
\newcommand{\eea}{\end{eqnarray}}

\newcommand{\tr}{\mbox{tr}}

\begin{document}

\pagestyle{plain}
\pagenumbering{arabic}

\maketitle

\begin{abstract}
This note quantifies the continuity properties of entanglement: how
much does entanglement vary if we change the entangled quantum state
just a little?  This question is studied for the pure state
entanglement of a bipartite system and for the entanglement of
formation of a bipartite system in a mixed state.
\end{abstract}


\begin{multicols}{2}[]
\narrowtext

%
%
Entanglement is a {\em resource} at the heart of quantum mechanics;
iron in the classical world's bronze age.  Entanglement plays a
crucial role in diverse quantum effects such as Bell inequalities
\cite{Peres93a}, quantum algorithms \cite{Shor97a,Grover97a}, quantum
teleportation \cite{Bennett93a}, and paradoxically is also responsible
for the emergence of a classical world out of the quantum
\cite{Zurek91a}.

%
%
To flesh out the notion that entanglement is a resource, various {\em
measures of entanglement} have been proposed to quantify the amount of
entanglement shared between two or more quantum systems.  For a pure
state of a two-party quantum system, Popescu and Rohrlich
\cite{Popescu97a} and Vidal \cite{Vidal98a} have shown that the
measure of entanglement is uniquely specified by certain natural
axioms\footnote{It is to be emphasized that these axioms are only
natural if one considers manipulations of large blocks of identically
entangled states\cite{Vidal98a}.}: it is given by the von~Neumann
entropy of the reduced density matrix associated with one of the
parties.  That is, if $|\psi\ra$ is the state of a composite system
with components $A$ and $B$, then the pure state entanglement of
$|\psi\ra$ is given by $E(\psi) = S(\rho_A) = S(\rho_B)$ where $\rho_A
\equiv \tr_B(|\psi\ra\la \psi|)$ and $\rho_B \equiv \tr_A(|\psi\ra \la
\psi|)$ are the reduced density matrices of system $A$ and system $B$,
respectively, and $S(\cdot)$ is the von~Neumann entropy.

%
%
The situation for mixed state entanglement is more complex, and a
plethora of measures have been developed (see
\cite{Bennett96a,Hill97a,Wootters98a,Vedral98a,Vidal98a} and other
references cited therein).  Perhaps the best understood of these
measures is the {\em entanglement of formation} studied in a series of
papers by Wootters and co-workers
\cite{Bennett96a,Hill97a,Wootters98a}.  For pure states the
entanglement of formation reduces to the von~Neumann entropy of the
reduced density matrix, as expected.  However, for mixed states the
entanglement of formation shows much more complex behaviour, a
behaviour that is not yet fully understood.

%
%
This note develops inequalities expressing continuity properties of
the pure state entanglement and the entanglement of formation.  We
begin with the simple arguments needed to prove such results for the
pure state entanglement.  This allows us to introduce some of the
tools needed for the more complex argument for entanglement of
formation, and also gives more stringent bounds than in the mixed
state case.  We will show that, up to constants, the bounds are
optimal with respect to the dimension of the underlying Hilbert space.
Furthermore, we explicitly show that the continuity bounds obtained
for the entanglement of formation and pure state entanglement are
stronger than those obtainable for other entanglement monotones,
continuing a line of thought initiated by Vidal \cite{Vidal98a}.  The
continuity bounds we obtain can be applied to analyze approximate
schemes for quantum communication protocols, quantum cloning, and
quantum communication complexity, work that will be presented
elsewhere.  The continuity of pure state entanglement has been
previously noted by the Horodecki family \cite{Horodecki99a}, although
explicit bounds on its variation were not given.

%
%
To understand how the entanglement between systems $A$ and $B$ varies
as we vary the density matrix for the combined system, we need to
introduce some {\em distance measures} on density matrices.  We will
make use of three closely related distance measures in our work, the
{\em trace distance}, the {\em fidelity}, and the {\em Bures
distance}.  To begin we need only the trace distance.  The trace
distance between density matrices $\rho$ and $\sigma$ is defined to be
$T(\rho,\sigma) \equiv \tr|\rho-\sigma|$.  It is easy to see that the
trace distance is a metric on the space of density matrices.
Furthermore, Ruskai \cite{Ruskai94a} has shown that the trace distance
is non-increasing under quantum operations.  That is, if ${\cal E}$ is
a trace-preserving quantum operation, then
\bea \label{eqtn:Ruskai}
T({\cal E}(\rho),{\cal E}(\sigma)) \leq T(\rho,\sigma), 
\eea
for all density matrices, $\rho$ and $\sigma$.  For our purposes, it
is especially important to note that this is true for the case where
${\cal E}$ is a partial trace operation, as the partial trace is a
trace-preserving quantum operation.

%
%
Fannes \cite{Fannes73a,Ohya93a} has proved a useful continuity
relation relating trace distance and entropy.  Fannes' inequality
states that for any density matrices $\rho$ and $\sigma$ such that
$T(\rho,\sigma) \leq 1/e$,
\bea \label{eqtn:Fannes}
|S(\rho)-S(\sigma)| \leq T(\rho,\sigma) \log(d)+\eta(T(\rho,\sigma)),
\eea
where $d$ is the dimension of the Hilbert space $\rho$ and $\sigma$
are defined on, $\eta(x) \equiv -x \log(x)$, and the base of
logarithms, here and throughout, is taken to be two.  It is useful to
note that $\eta(x)$ is increasing for $0 \leq x \leq 1/e$.  The
restriction on~(\ref{eqtn:Fannes}) that $T(\rho,\sigma) \leq 1/e$ may
be lifted to give
\bea \label{eqtn:Fannes_2}
|S(\rho)-S(\sigma)| \leq T(\rho,\sigma) \log(d)+\frac{\log(e)}{e}.
\eea

%
%
Ruskai's result~(\ref{eqtn:Ruskai}) can be combined with Fannes'
inequality~(\ref{eqtn:Fannes}) to obtain the desired continuity
relation for pure state entanglement.  Suppose $|\psi\ra$ and
$|\phi\ra$ are pure states of a composite quantum system with
components $A$ and $B$, and that system $A$ has dimension $d$.  Let
$\rho_A$ and $\sigma_A$ be the corresponding reduced density matrices
for system $A$.  Applying Fannes' inequality~(\ref{eqtn:Fannes}) gives
\bea
|E(\psi)-E(\phi)| & = & |S(\rho_A)-S(\sigma_A)| \\
	& \leq & T(\rho_A,\sigma_A)
\log(d)+\eta(T(\rho_A,\sigma_A)).
\eea
Recalling that $\eta(x)$ is monotonically increasing for $0 \leq x
\leq 1/e$, and using~(\ref{eqtn:Ruskai}) to deduce that
$T(\rho_A,\sigma_A) \leq T(\psi,\phi)$, we obtain
\bea \label{eqtn:pure_continuity}
|E(\psi)-E(\phi)| \leq T(\psi,\phi) \log d + \eta(T(\psi,\phi)),
\eea
provided $T(\psi,\phi) \leq 1/e$.  This is the desired continuity
relationship for the pure state entanglement.
Using~(\ref{eqtn:Fannes_2}) we may lift the restriction $T(\psi,\phi)
\leq 1/e$ to give the bound
\bea \label{eqtn:un_pure_cont}
|E(\psi)-E(\phi)| \leq T(\psi,\phi) \log d + \frac{\log(e)}{e}.
\eea

%
%
We now generalize these pure state results to apply to the
entanglement of formation.  Our strategy for proving a continuity
bound for the entanglement of formation involves three ingredients in
addition to those used in the proof of the bound for the pure state
entanglement~(\ref{eqtn:pure_continuity}): Uhlmann's formula for the
{\em fidelity} \cite{Uhlmann76a}, the {\em Bures distance}
\cite{Bures69a}, and the {\em remote-control} view of entanglement
\cite{DiVincenzo98b}.

%
%
First, however, we must define the entanglement of formation.  For a
density matrix $\rho$ of a composite system $AB$ the entanglement of
formation is defined by \cite{Wootters98a}
\bea \label{eqtn:min}
E(\rho) \equiv \min \sum_m p_m S(\rho_{A,m}),
\eea 
where the minimization is over all ensembles $\{ p_m,|AB_m\ra \}$
generating the state $\rho$, that is, $\rho = \sum_m p_m |AB_m\ra \la
AB_m|$, and $\rho_{A,m} \equiv \tr_B(|AB_m\ra \la AB_m|)$.  In
practice, evaluating this expression seems to be very difficult; all
that is known is an ingenious expression for the entanglement of
formation of a pair of qubits found by Wootters \cite{Wootters98a},
building on earlier work by Hill and Wootters \cite{Hill97a}.

%
%
The first ingredient needed to prove the continuity bound for the
entanglement of formation is the {\em fidelity}, a measure of distance
between two density matrices distinct from but closely related to the
trace distance.  The fidelity between density matrices $\rho$ and
$\sigma$ is defined to be $F(\rho,\sigma) \equiv \tr \sqrt{\rho^{\frac
12} \sigma \rho^{\frac 12}}$.  For pure states $|\psi\ra$ and
$|\phi\ra$ the fidelity reduces to the overlap between the states,
$F(\psi,\phi) = |\la \psi| \phi\ra|$.  The fidelity is not a metric,
however it does possess many useful properties as a measure of
distance, and is closely related to the trace distance and will be
used in the definition of the Bures distance
\cite{Fuchs99a,Fuchs96a,Nielsen98d}.  Uhlmann
\cite{Uhlmann76a,Jozsa94c} has found a useful expression for the
fidelity relying on the following construction.  Suppose $\rho$ and
$\sigma$ are quantum states of a $d$-dimensional quantum system.  We
label the system $Q$ for convenience.  Introduce an additional
``reference'' system $R$ \cite{Schumacher96a} of any fixed
dimensionality that is at least as great as $d$.  Uhlmann's expression
for the fidelity is
\bea \label{eqtn:Uhlmann}
F(\rho,\sigma) = \max | \la \rho | \sigma \ra | = \max F(|\rho\ra,|\sigma\ra), 
\eea 
where the maximization is performed over all pure states $|\rho\ra$
and $|\sigma\ra$ of the joint system $RQ$ such that $\tr_R(|\rho\ra
\la \rho|) = \rho$ and $\tr_R(|\sigma\ra \la \sigma|) = \sigma$; such
states are known as {\em purifications} of $\rho$ and $\sigma$.

%
%
As our second ingredient, we introduce the {\em Bures distance}
\cite{Bures69a} between density matrices $\rho$ and $\sigma$,
\bea
D(\rho,\sigma) \equiv 2 \sqrt{1-F(\rho,\sigma)}.
\eea
$D(\cdot,\cdot)$ is easily shown to be a metric on the space of
density matrices.  We have chosen the overall normalization factor of
$2$ out the front so the Bures distance $D(\cdot,\cdot)$ agrees with
the trace distance for pure states; other authors often use different
normalizations.

%
%
The third ingredient we need is the elegant remote-control view of
entanglement \cite{DiVincenzo98b}.  Suppose $\rho$ is some joint state
of a composite system $AB$, where $A$ is $d$-dimensional and $B$ is
$d'$-dimensional.  Introduce a $d^2d'^2$-dimensional reference system
$R$ which purifies those systems into a pure state $|\rho\ra$.  Let
$\{ p_m, |AB_m\ra \}$ be the ensemble of states achieving the minimum
in~(\ref{eqtn:min}).  Uhlmann \cite{Uhlmann97b} has shown that it is
possible to achieve this minimum using an ensemble containing at most
$d^2 d'^2$ ensemble elements.  A result of Hughston, Jozsa, and
Wootters \cite{Hughston93a} implies that by performing an measurement
on $R$ with respect to an appropriate orthonormal basis $|m\ra$, the
corresponding posterior states of $AB$ will be $|AB_m\ra$, with
probability $p_m$.  Elementary calculation show that after the
measurement we have $S(R') = H(p_m)$ and $S(AR') = H(p_m)+E(\rho)$,
where $S(R')$ denotes the von~Neumann entropy of $R$ after the
measurement, and similarly for $S(AR')$.  Combining these observations
results in the very useful expression
\bea
E(\rho) = S(AR')-S(R').
\eea
If instead a measurement had been performed in some other orthonormal
basis $|m'\ra$ then we would have had
\bea \label{eqtn:remote_inequality}
E(\rho) \leq S(AR')-S(R').
\eea

%
%
Let us now proceed to the proof of the continuity relation for the
entanglement of formation.  Let $\rho_{AB}$ and $\sigma_{AB}$ be two
density matrices of the system $AB$, where $A$ has $d$ dimensions, and
$B$ has $d'$ dimensions.  Introduce a $d^2 d'^2$ dimensional reference
system $R$.  By Uhlmann's formula~(\ref{eqtn:Uhlmann}) there exist
purifications $|\rho \ra$ and $|\sigma \ra$ of $\rho_{AB}$ and
$\sigma_{AB}$ to the system $ABR$ such that
\bea \label{eqtn:Uhlmann_2}
F(\rho_{AB},\sigma_{AB}) = F(|\rho \ra,|\sigma \ra).
\eea
Suppose we measure system $R$ in a basis chosen such that
$E(\sigma_{AB}) = S(\sigma_{AR}')-S(\sigma_R')$, where the primes
denote density matrices after the measurement, and the initial state
was $|\sigma\ra$.  Performing the same measurement with initial state
$|\rho\ra$ we see from~(\ref{eqtn:remote_inequality}) that
$E(\rho_{AB}) \leq S(\rho_{AR}')-S(\rho_R')$.  Taking the difference
of these equations yields
\bea
E(\rho_{AB})-E(\sigma_{AB}) & \leq & 
S(\rho_{AR}')- S(\sigma_{AR}') \nonumber \\
& & +S(\sigma_R')- S(\rho_R').
\eea
Applying Fannes' inequality~(\ref{eqtn:Fannes}) twice on the right
hand side gives
\bea
& & E(\rho_{AB})-E(\sigma_{AB}) \leq \nonumber \\
& & \log(d^3d'^2) T(\rho_{AR}',\sigma_{AR}')+ \eta(T(\rho_{AR}',\sigma_{AR}'))
	\nonumber \\ 
\label{eqtn:continuity_1} & & + \log(d^2d'^2)T(\rho_R',\sigma_R') +\eta(T(\rho_R',\sigma_R')).
\eea
By~(\ref{eqtn:Ruskai}) we have
\bea & & 
T(\rho_R',\sigma_R') \leq T(\rho_{AR}',\sigma_{AR}') \leq
T(\rho_{ABR}',\sigma_{ABR}') \leq T(|\rho\ra,|\sigma\ra). \nonumber \\
& &
\eea
Recall that $T(|\rho\ra,|\sigma\ra) = D(|\rho\ra,|\sigma\ra)$.
Together with the previous equation this fact
and~(\ref{eqtn:Uhlmann_2}) gives
\bea
T(\rho_R',\sigma_R') \leq T(\rho_{AR}',\sigma_{AR}')  \leq
D(\rho_{AB},\sigma_{AB}).
\eea
Combining this equation with~(\ref{eqtn:continuity_1}) gives
\bea
E(\rho_{AB})-E(\sigma_{AB}) & \leq & (5\log(d)+4\log(d')) D(\rho_{AB},\sigma_{AB})
	\nonumber \\  \label{eqtn:continuity}
	& & + 2\eta(D(\rho_{AB},\sigma_{AB})),
\eea
provided $D(\rho_{AB},\sigma_{AB}) \leq 1/e$.  This is the desired
continuity equation for the entanglement of formation.  Of course, the
role of $A$ and $B$ may be interchanged in this expression; clearly
the strongest inequality is obtained by labeling the systems such that
$d \leq d'$.  For many purposes it is sufficient to replace the
logarithmic terms in the right hand side by $9\log(\max(d,d'))$.

%
%
The restriction $D(\rho_{AB},\sigma_{AB}) \leq 1/e$
on~(\ref{eqtn:continuity}) may be lifted in a manner similar to that
for the continuity bound for pure state entanglement.  Doing so gives 
\bea \label{eqtn:unrestricted_continuity}
 E(\rho_{AB})-E(\sigma_{AB}) & \leq & (5 \log(d)+4 \log(d'))
D(\rho_{AB},\sigma_{AB}) \nonumber \\
 & & + 2\log(e)/e.
\eea

%
%
For applications to communication in which large blocks of
entanglement are used and $d$ becomes large it is desirable to
understand how close to optimal (with respect to $d$) the
bounds~(\ref{eqtn:continuity})
and~(\ref{eqtn:unrestricted_continuity}) are.  Understanding this is
essentially the problem of understanding how close to optimal Fannes'
inequality is.  Let $\epsilon > 0$ be given, and for a $d$-dimensional
Hilbert space with orthonormal basis $|1\ra,\ldots,|d\ra$ define $\rho
\equiv \epsilon d |1\ra\la 1| + (1/d-\epsilon ) I$.  For small
$\epsilon$ this is a density matrix close to the completely mixed
state $I/d$.  We will analyze the difference in entropies between
$\rho$ and $I/d$.  From the general bound \cite{Wehrl78a} $S(\sum_i
p_i \rho_i) \leq H(p_i) +\sum_i p_i S(\rho_i)$ where $H(\cdot)$ is the
Shannon entropy, we obtain
\bea
S(\rho) \leq (1-\epsilon d) \log(d)+H(\epsilon d,1-\epsilon d).
\eea
Thus
\bea
S(I/d)-S(\rho) & \geq & \epsilon d \log(d) - H(\epsilon d,1-\epsilon
d)  \\
& \geq & \epsilon d \log(d) -1.
\eea
A simple calculation shows that $T(I/d,\rho) = 2(d-1)\epsilon$.  It
follows easily that
\bea
S(I/d)-S(\rho) \geq \frac{T(I/d,\rho) \log(d)}{2} -1.
\eea
This implies that the logarithmic behaviour (with $d$) expressed
in~(\ref{eqtn:Fannes}) and thus
in~(\ref{eqtn:pure_continuity}),~(\ref{eqtn:un_pure_cont}),~(\ref{eqtn:continuity})
and~(\ref{eqtn:unrestricted_continuity}) cannot be improved beyond a
constant factor.

%
%
Vidal \cite{Vidal98a} has emphasized the importance of continuity to
the result of Popescu and Rohrlich \cite{Popescu97a} (see also
\cite{Horodecki99a}).  Popescu and Rohrlich argue that any measure of
bipartite pure state entanglement that is (a) additive, and (b)
non-increasing under local operations and classical communication, is
necessarily proportional to the von~Neumann entropy of the reduced
density matrix of the pure state.  Vidal pointed out some hidden
assumptions in this argument by explicitly constructing examples of
entanglement measures that are additive and do not increase under
local operations and classical communication, yet are not proportional
to the von~Neumann entropy.  For example, a function with the required
properties is $\tilde E(\psi) \equiv -\log(\tr(\rho^2))$, which is
manifestly different from the von~Neumann entropy.  Vidal points out
that the key property lacking in such a measure is sufficiently strong
{\em continuity} properties.  The framework of the present note
provides a useful opportunity to elaborate.  Suppose $\tilde E(\rho)$
is any additive measure of entanglement that does not increase under
local operations and classical communication.  We will show that
$\tilde E(\rho)$ cannot satisfy a continuity property as strong
as~(\ref{eqtn:continuity}) unless for pure states it is proportional
to the von~Neumann entropy.  Indeed for any constants $C$ and $D$ we
will show that a continuity property as strong as
\bea \label{eqtn:false}
|\tilde E(\rho)-\tilde E(\sigma)| \leq C \log(d)D(\rho,\sigma)+ D
\eea
implies the $\tilde E$ is proportional to the von~Neumann entropy of
the reduced density matrix, where $d$ is the maximum of the dimensions
of system $A$ and system $B$.  Thus the von~Neumann entropy is in some
sense the ``most continuous'' measure of entanglement, satisfying a
stronger bound on its variation than any other prospective measure of
entanglement.  Suppose~(\ref{eqtn:false}) holds. Let $\epsilon > 0$ be
given.  Then for sufficiently large $n$ entanglement dilution
\cite{Bennett96c} allows us to convert from $n(S(\rho)+\epsilon)$ Bell
states into a state $\sigma$ that satisfies $D(|\psi\ra^{\otimes
n},\sigma) \leq \epsilon$, using local operations and classical
communication.  Then if~(\ref{eqtn:false}) holds,
\bea
n \tilde E(\psi) = \tilde E(\psi^{\otimes n}) & \leq & \tilde
E(\sigma)+C\log(d^n)\epsilon +D,
\eea
since $D(|\psi\ra^{\otimes n},\sigma) \leq \epsilon$.  The
non-increase of $\tilde E$ under local operations and classical
communication implies that $\tilde E(\sigma) \leq
n(S(\rho)+\epsilon)k$, where $k$ is the entanglement associated with a
single Bell pair according to the measure $\tilde E$.  Thus
\bea
n\tilde E(\psi) \leq n(S(\rho)+\epsilon)k+Cn\log(d)\epsilon+D.
\eea
Dividing by $n$ and letting $\epsilon \rightarrow 0, n \rightarrow
\infty$ gives $\tilde E(\psi) \leq k S(\rho)$.  Similarly, for any
$\epsilon >0$ and sufficiently large $n$, entanglement concentration
\cite{Bennett96c} allows us to convert $n$ copies of $|\psi\ra$ into a
state $\tau$ satisfying $D(|\beta\ra^{\otimes
n(S(\rho)-\epsilon)},\tau) < \epsilon$, where $|\beta\ra$ is a Bell
state.  By~(\ref{eqtn:false}) we have
\bea
n(S(\rho)-\epsilon) k \leq \tilde E(\tau)+Cn \log(d) \epsilon + D.
\eea
But $\tilde E(\tau) \leq n\tilde E(\psi)$ since $\tilde E$ is
non-increasing under local operations and classical communication.
Thus
\bea
n(S(\rho)-\epsilon) k & = & \tilde E(\beta^{\otimes n(S(\rho)-\epsilon)})
  \\
& \leq & n \tilde E(\psi)+Cn \log(d) \epsilon + D.
\eea
Dividing by $n$ and letting $\epsilon \rightarrow 0, n \rightarrow
\infty$ gives $kS(\rho) \leq \tilde E(\psi)$.

Combining the results of the last paragraph, we see that the
properties of being additive, non-increasing under local operations
and classical communication, and satisfying~(\ref{eqtn:false}) for
some $C$ and $D$, imply that $\tilde E(\psi) = k S(\rho)$.  Thus
measures of pure state entanglement such as $\tilde E(\psi) =
-\log(\tr(\rho^2))$ which are not proportional to $S(\rho)$ must
satisfy weaker continuity relations than~(\ref{eqtn:false}).

%
%
Finally, it should be mentioned that the
bounds~(\ref{eqtn:continuity})
and~(\ref{eqtn:unrestricted_continuity}) apply only to the
entanglement of formation as defined in~(\ref{eqtn:min}).  As
discussed by Wootters \cite{Wootters98a}, the interpretation of the
entanglement of formation as defined in~(\ref{eqtn:min}) may be
somewhat problematic.  The basic problem is that one would like to
interpret the entanglement of formation as a measure of the {\em
resources} --- Bell pairs --- that must be shared between Alice and
Bob in order to {\em create} $\rho$.  That is, if Alice and Bob are
provided with $nE(\rho)$ Bell pairs, then by local operations and
classical communication they can convert them with high fidelity into
$n$ copies of $\rho$, in the limit that $n$ is large.  As Wootters
discussed, the only impediment to this interpretation is the question
of whether or not the entanglement of formation is {\em additive}.
That is, is it true that $E(\rho^{\otimes n}) = n E(\rho)?$ If this is
not true, then it suggests a revised definition for the entanglement
of formation, the operational entanglement of formation, as
$E_{op}(\rho) \equiv \limsup_{n \rightarrow \infty} E(\rho^{\otimes
n})/n$.  $E_{op}(\rho)$ quantifies the resources needed for Alice and
Bob to create $\rho$, in the sense described above.  Unfortunately,
the reasoning used in the derivation of the continuity bounds on the
entanglement of formation~(\ref{eqtn:continuity})
and~(\ref{eqtn:unrestricted_continuity}) does not go through for
$E_{op}(\rho)$.  It is an interesting open problem to determine such
bounds for the operational entanglement of formation, and may yield
insight into the question of whether or not the entanglement of
formation is additive.

%
%
We have obtained a continuity relation for bipartite pure state
entanglement and the entanglement of formation.  This relation bounds
the variation of the entanglement $E(\rho)$ between two systems, $A$
and $B$, as the state of the joint system $\rho$ is varied.  The bound
obtained exhibits the best possible behaviour with respect to the
dimension $d$ of the underlying Hilbert space, to within constant
factors, and is stronger than the continuity bounds that may be
obtained for other potential measures of entanglement.  Further
applications to quantum communication protocols, quantum cloning, and
quantum communication complexity will be reported elsewhere.

\section*{Acknowledgments}
This research was supported by DARPA through the Quantum Information
and Computing Institute (QUIC) administered through the ARO, and by
the California Institute of Technology through a Tolman Fellowship.

\end{multicols}

\end{document}